# A Framework for Enabling Safe and Resilient Food Factories for Public Feeding Programs

Nataraj Kuntagod, Sanjay Podder, Satya Sai Srinivas Abbabathula, Venkatesh Subramanian, Giju Mathew, Suresh Kumar Mani
Accenture, Bangalore, India.
Email: {nataraj.s.kuntagod, sanjay.podder, satya.sai.srinivas, venkatesh.subramania, giju.mathew, suresh.kumar.mani}@accenture.com

*Abstract*—Public feeding programs continue to be a major source of nutrition to a large part of the population across the world. Any disruption to these activities, like the one during the Covid-19 pandemic, can lead to adverse health outcomes, especially among children. Policymakers and other stakeholders must balance the need for continuing the feeding programs while ensuring the health and safety of workers engaged in the operations. This has led to several innovations that leverage advanced technologies like AI and IOT to monitor the health and safety of workers and ensure hygienic operations. However, there are practical challenges in its implementation on a large scale.

This paper presents an implementation framework to build resilient public feeding programs using a combination of intelligent technologies. The framework is a result of piloting the technology solution at a facility run as part of a large mid-day meal feeding program in India. Using existing resources like CCTV cameras and new technologies like AI and IOT, hygiene and safety compliance anomalies can be detected and reported in a resource-efficient manner. It will guide stakeholders running public feeding programs as they seek to restart suspended operations and build systems that better adapt to future crises.

## I. INTRODUCTION

Access to nutritious food is a critical social determinant of health. For many people, especially children, a public feeding program is the primary source of a proper meal. This is true for low-income households across the world, as access to free meals also serves as a form of financial aid. The restrictions imposed as a result of the Covid-19 pandemic exposed the fragility of these programs and the severe impact such disruptions can have on citizens' health and well-being. The loss of income and livelihoods, disruptions in social protection programs, and school closures have all negatively impacted food security and nutrition [1]. For example, in the US, during the initial lockdown in April 2020, more than one in five households, and two in five households with mothers with children 12 and under, were food insecure [2]. The World Food Programme estimates that in April 2020, more than 369 million children were missing out on meals at schools globally at the peak of school closures. Of this, 47 percent were girls [3]. The lack of access to nutrition during disruptions could further deteriorate health outcomes with weakened immune systems and the risk of disease. In some countries, school feeding programs contribute to two-thirds of children's daily nutritional needs [4]. In countries like India, where child undernutrition has increased between 2015 and 2019, in part due to the economic slowdown, the pandemic is only going to make it worse [5].

The challenge for stakeholders such as the government and non-profits is to ensure continuity of public-feeding programs during a crisis without compromising the health and safety of workers who run these operations. While technology solutions offer hope, they need to account for a food-factory's unique requirements and adhere to individual privacy rights. This paper presents an actionable framework that can be applied to build resilient public-feeding programs, based on our experience of piloting this solution in one of the world's largest school meal program.

## II. RELATED WORK

Recent studies have looked at the impact of the Covid-19 pandemic on nutrition, particularly for vulnerable groups. Studies have suggested actions that policymakers need to undertake to reduce stress on already strained healthcare systems [6]. Multiple studies highlight the adverse impact on school feeding programs and the need to extend emergency benefits to address food insecurity among households with children [7] [8].

The very nature of school feeding programs that are feasible only in a physical learning setting is questioned. The limitations of these programs and ensuring children's nutritional needs are met all year round call for a more sustainable school feeding program [9]. Studies have evaluated government responses to disruptions in feeding programs during school closures and called for better emergency planning to help address food insecurity, limit disease transmission and prevent health disparities [10] [11]. Research has also highlighted the need to enhance government capacity and improve relationships with non-profits to respond better to a food crisis [12].

Research appears to be exploratory in nature about using intelligent technologies to ensure the safety and health of workers while running public feeding programs during a pandemic. However, multiple studies have analyzed the use of technology to encourage hand hygiene. Automated hand hygiene monitoring systems facilitate data collection, which can then nudge people towards a behavior change [13] [14]. Deep learning-based social distance monitoring models have been developed to track and identify humans in video sequences. They help ensure compliance with mandated social distance requirements between people in closed areas [15]. Others have proposed AI-based real-time social distancing detection and warning systems that emits an audio-visual warning if a violation is detected without targeting a specific individual [16]. IoT-based systems that support indoor safety monitoring, including mask detection and social distancing check, have also been experimented with [17]. The use of mobile robots with sensors and wearable magnetic field-based proximity sensing systems to monitor social distancing requirements in crowded scenarios have been piloted [18] [19].

Several AI and IOT based solutions exist that focus on various aspects of compliance monitoring around safety and hygiene. However, there are considerations around the cost and complexity of these solutions that need to be addressed. The need is for a solution that brings together these various capabilities in a cost-effective manner and meets the unique needs of an indoor factory environment.

### III. Key Challenges In Building Resilient Public Feeding Programs

The susceptibility of public feeding programs to potential disruptions is evident from the experience during the Covid-19 pandemic. In response, while building up the resilient public feeding programs in food-factories, through automated monitoring systems, several practical implementation challenges have emerged.

*Constrained resources for deploying new digital technologies:* The social enterprises that run these large feeding programs are funded by the government and private donors, with a mandate to utilize the funds mostly for public feeding of vulnerable population while keeping the overheads low. This has resulted in low penetration of modern digital infrastructure in the operations. For example, Akshaya Patra, the world's largest NGO-run school feeding programme, utilizes 92% of funds on feeding the children and just 8% on management & infrastructure.

*Operational diversity – small and large format food factories*: Large public feeding programs like Akshaya Patra have large centralized factories generating 200,000 meals/day and small format ones producing 40,000 meals/day. Based on the region, the meals produced are also different. For example, North Indian and South Indian meals that have different production process. The standard operating procedures (SOP) will have variations across these factory/meal types, and as we learn more about the pandemic, the SOPs will also evolve.

*Preserving individual privacy rights:* Technologies like AI/IOT can give unprecedented insights into the operational anomalies at an individual staff level. The challenge is to build a solution that can give granular actionable insights of SOP violations, while still preserving privacy.

### IV. Solution Approach

The three broad categories for compliance monitoring are

1. Safety attire: facemask, hairnet, gloves, apron
2. Cleaning activity: mopping, hand sanitization/washing, and vessel sterilization
3. Social distancing and indoor contact tracing

Compliance activities are further classified into those that require immediate attention or can tolerate delays in reporting without adversely impacting the operational safety [Table 1]. This classification helps to distribute the anomaly detection processing on the cloud or have specialized "on-premise" edge processing hardware/software. The approach is to limit the on-premise deployments to reduce the complexity and cost of maintaining the edge systems.

*Table 1: SOP Priorities*

| Violations | | Priority |
|---|---|---|
| SOP for safety attire and cleaning activities | Face mask | Delay Tolerant |
| | Hairnet | Delay Tolerant |
| | Sterilization | Delay Tolerant |
| | Mopping | Delay Tolerant |
| | Handwash | Immediate |
| Social Distancing | | Delay Tolerant |
| Contact tracing | | Immediate |

The proposed framework's [Fig. 1] loosely coupled layers allows for independent evolution of each layer and provide separate deployment options. Most of the functions run in "serverless" mode to keep the cloud cost low.

*Anomaly detection*: Uses existing video feeds from CCTV cameras for object detection and activity recognition [20]. Video feeds are batch processed in the cloud instead of continuous monitoring. The batch processing is also synchronized with the food factory operational schedule to prevent unnecessary footages being analyzed. Local processing is done only for those that require immediate reporting and is kept to a minimum.

Realtime Indoor location Technology based on Bluetooth-Low Energy (BLE) provides alerts for social distancing violations and provides indoor contact tracing in case of an infection detection [21]. The cloud-based service is asset light and only has BT gateways deployed in the premises with BT badges that are carried by the staff. This layer allows more such external anomaly detection services to be incorporated in future and has pay-per-use model keeping the operating expenses low.

Anomaly detection services do not store Personally Identifiable Information (PII). The result is a robust privacy compliant system that identifies spaces that have compliance issues, and flag those areas as red, amber, or green, based on the number of violations. However, since the solution does not identify the person this can result in the same person's violation repeatedly appearing as new violation. To avoid the event flooding caused by duplicate detection the solution has an event similarity score logic to reconcile the same person's violation appearing repeatedly.

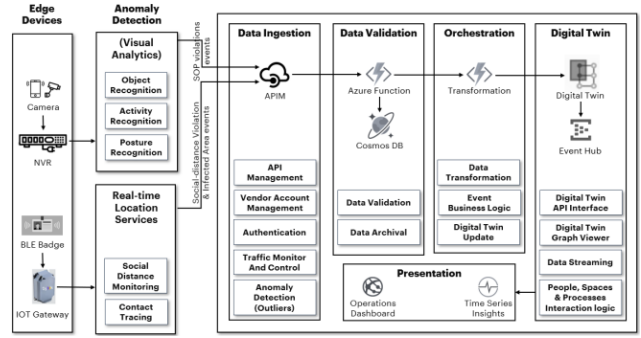

*Fig. 1: Reference Framework*

*Data Ingestion*: Azure API Management (APIM) service is used to ingest the standardized anomaly events generated from multiple external services. APIM enables quick onboarding of vendor anomaly detection solutions, while providing common functions like authentication, routing, rate limiting, billing, monitoring, policies and security.

*Data validation and Orchestration*: Generates higher order contextual information out of the anomaly events from multiple systems and feeds into the digital twin of the factory

*Digital Twin* is the core of our framework and helps in managing the operational diversity of small and large format food factory. The entire public feeding program is modelled as multiple hierarchical areas(spaces) mapping to the real physical layout, people who are working in that space and process related to those spaces [*Fig. 2*]. The relationships are expressed as an ontology in Azure Digital Twin definition language. Spaces in food factory have different policies for SOP and these are represented in central repository of the Digital twin. A template public feeding program model is used as the base to derive the specific models for centralized and small format distributed food factory, which further is customized based on the meal type (North / South Indian plan). The digital twin maps incoming events from the physical domain to the correct virtual space, process, and/or people in the digital graph. Further, the same data is also visible by means of a web dashboard to the operation staff. This synthesis of the data results in a cognitively simple interface for supervisory staff to respond to events in a timely manner. The non-compliance violation, be it a person-specific (hairnet, face mask, social distancing) or space-specific (mopping, sanitization, infected area), is mapped back to the space where that violation happened, and depending upon the number of violations, each space is flagged red, amber and green [Fig. 3]. The loose coupling between spaces and processes allow us to change the space specific processes dynamically. Similarly, the loose coupling between spaces and people allows us to shuffle the staff across spaces.

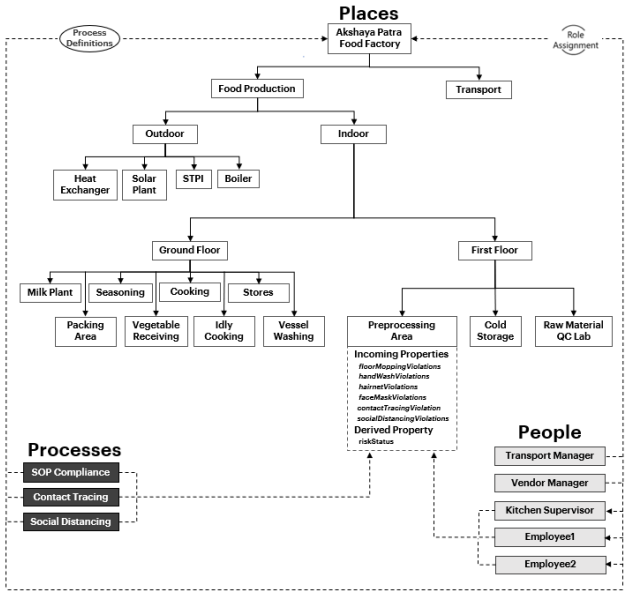

*Fig. 2: Digital Twin Hierarchical View*

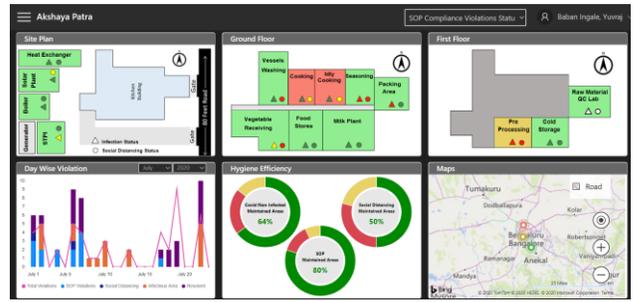

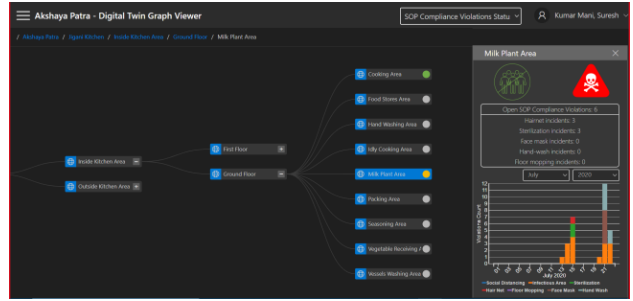

*Fig. 3: Integrated Dashboard & Digital Twin Viewer*

## V. PILOT RESULTS AND DISCUSSION

The solution was deployed at the Akshaya Patra's food factory at Jigani, Karnataka, India, manned by 180 operations staff, generating 40,000 meals/day. The total floor area is 21,000 square feet comprising of 16 different areas like packing, food stores, milk plan, hand washing, preprocessing, cold storage, seasoning, vegetable receiving, cooking etc. where compliance was monitored over a period of 21 days.

Since the factory was in reduced capacity operation due to COVID, we were able to experiment in places that were not in operation by injecting SOP compliance violations and checking the system's sensitivity. The solution identified 129 SOP compliance violations over this period. The system sensitivity and specificity for hygiene and safety SOP compliance was 97% and 99%, while for tracking and tracing, it was 100% & 92% respectively. The field staff were interviewed during this period to gauge the efficacy of the system, and the feedback has been positive, as the system and the processes are more focused towards group behavior change than at an individual level, while still giving enough actionable insights for corrective actions.

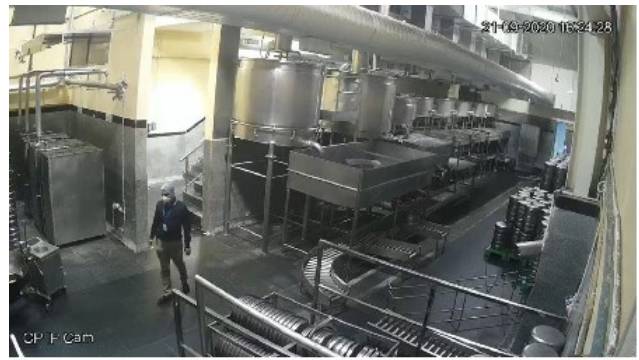

*Fig. 4: Cooking Area*

The off-the-shelf visual analytics solution base object detection and activity detection was retrained for each space with 15 hours of video footage, as the initial accuracy was low due to the visual clutter and varied lighting conditions and the CCTV cameras were at a higher level, looking down on each space, and not at eye level [Fig. 4]. The accuracy improved only after retraining. Availability of video footages for each space spread across the entire day is a prerequisite for rapid deployment of the system across multiple food factory formats.

Automatic activity completion detection like 100% mopping of an area needs further research. The generic activity detection service can accurately identify the activity (E.g. mopping) but cannot be extended to predict the percentage completion in an area. However, time in activity was used as a proxy for estimating the percentage completion.

In a high-touch closed-room environments like food factories, the areas where the infected person has moved around is marked as "at risk" and all the people who have visited the risk areas within a threshold period of time, are also marked as a potential risk. This feature was introduced in the real time location service, which focused on direct person-to-person contact, and not indirect contact. This feature allowed identification of specific indoor spaces for sanitization, and not shut down the entire factory, helping Akshaya Patra to have lesser disruption in their time sensitive meal preparation activities, ensuring timely supply of nutritious and safe food to the children every day.

As Akshaya Patra refines the SOP and bring out automated mitigation procedures, the digital twin's capability to call user defined functions for initiating the automated mitigation procedures with other systems in Akshaya Patra can be explored in future.

## VI. Conclusion

The proposed framework is built on a combination of advanced technologies and considers the practical challenges in implementation. It offers stakeholders of public feeding programs to build resilience and be better prepared for the next crisis – which may be different from the one we are experiencing now. While the solution is technology-driven, the end-objective is drive behavior change and make safety and hygiene practices stronger. The solution goes beyond the immediate need for continuity of operations and helps ensure that the health and nutritional needs of the most vulnerable are protected. The framework brings together the various partners in the ecosystem and leverages the power of technology for the greater good.

## VII. Acknowledgement

We would like to thank Yogesh Mallaiah and Rakshit Debjyoti for their contribution in the development of the solution and field pilot.## References

[1] HLPE. 2020. Impacts of COVID-19 on food security and nutrition: developing effective policy responses to address the hunger and malnutrition pandemic. Rome. https://doi.org/10.4060/cb1000en

[2] Bauer, L. (2020, May 12). The COVID-19 crisis has already left too many children hungry in America. Retrieved January 06, 2021, from https://www.brookings.edu/blog/up-front/2020/05/06/the-covid-19-crisis-has-already-left-too-many-children-hungry-in-america/

[3] Global Monitoring of School Meals During COVID-19 School Closures, https://cdn.wfp.org/2020/school-feeding-map/?_ga=2.51872297.1444362342.1609956343-1105941433.1609956343

[4] Dunn, C. G., Kenney, E., Fleischhacker, S. E., & Bleich, S. N. (2020). Feeding Low-Income Children during the Covid-19 Pandemic. New England Journal of Medicine, 382(18). doi:10.1056/nejmp2005638

[5] International Institute for Population Sciences. (2020). National Family Health Survey (NFHS-5), 2019-20: India. Mumbai, India: International Institute for Population Sciences. http://rchiips.org/NFHS/NFHS-5_FCTS/NFHS-5%20State%20Factsheet%20Compendium_Phase-I.pdf

[6] Headey, Derek D. & Ruel, Marie T., 2020. "The COVID-19 nutrition crisis: What to expect and how to protect," IFPRI book chapters, in: COVID-19 and global food security, chapter 8, pages 38-41, International Food Policy Research Institute (IFPRI).

[7] Amorim, Ana Laura Benevenuto de, Ribeiro Junior, José Raimundo Sousa, & Bandoni, Daniel Henrique. (2020). National School Feeding Program: strategies to address food insecurity during and after COVID-19. Public Administration Journal, 54(4), 1134-1145. Epub August 28, 2020. https://doi.org/10.1590/0034-761220200349

[8] Bauer, L. (2020, May 12). The COVID-19 crisis has already left too many children hungry in America. Retrieved January 06, 2021, from https://www.brookings.edu/blog/up-front/2020/05/06/the-covid-19-crisis-has-already-left-too-many-children-hungry-in-america/

[9] Amolegbe K. B. (2020). Hungry birds do not sing: Coronavirus and the school feeding program. World development, 136, 105169. https://doi.org/10.1016/j.worlddev.2020.105169

[10] McLoughlin, G. M., Fleischhacker, S., Hecht, A. A., McGuirt, J., Vega, C., Read, M., ... & Dunn, C. G. (2020). Feeding Students During COVID-19—Related School Closures: A Nationwide Assessment of Initial Responses. Journal of nutrition education and behavior, 52(12), 1120-1130.

[11] Correa, E. N., NEVES, J. D., SOUZA, L. D. D., LORINTINO, C. D. S., & Porrua, P. (2020). School feeding in Covid-19 times: mapping of public policy execution strategies by state administration. Revista de Nutrição, 33.

[12] Barker, M., & Russell, J. (2020). Feeding the food insecure in Britain: learning from the 2020 COVID-19 crisis. Food Security, 12(4), 865-870.

[13] Cawthorne, K. R., & Cooke, R. P. (2020). Innovative technologies for hand hygiene monitoring are urgently needed in the fight against COVID-19. Journal of Hospital Infection, 105(2), 362-363.

[14] Hess, O. C., Armstrong-Novak, J. D., Doll, M., Cooper, K., Bailey, P., Godbout, E., ... & Bearman, G. (2020). The impact of coronavirus disease 2019 (COVID-19) on provider use of electronic hand hygiene monitoring technology. Infection Control & Hospital Epidemiology, 1-3.

[15] Ahmed, I., Ahmad, M., Rodrigues, J. J., Jeon, G., & Din, S. (2020). A deep learning-based social distance monitoring framework for COVID-19. Sustainable Cities and Society, 102571.

[16] Yang, D., Yurtsever, E., Renganathan, V., Redmill, K. A., & Özgüner, Ü. (2020). A Vision-based Social Distance and Critical Density Detection System for COVID-19. arXiv preprint arXiv:2007.03578.

[17] Petrović, N., & Kocić, Đ. (2020). IoT-based System for COVID-19 Indoor Safety Monitoring. preprint), IcETRAN, 2020, 1-6.

[18] Sathyamoorthy, A. J., Patel, U., Savle, Y. A., Paul, M., & Manocha, D. (2020). COVID-robot: Monitoring social distancing constraints in crowded scenarios. arXiv preprint arXiv:2008.06585.

[19] Bian, S., Zhou, B., Bello, H., & Lukowicz, P. (2020, September). A wearable magnetic field based proximity sensing system for monitoring COVID-19 social distancing. In Proceedings of the 2020 International Symposium on Wearable Computers (pp. 22-26).

[20] Wobot Intelligence - https://wobot.ai/

[21] Syook - https://syook.com/